\documentclass[aps,amsmath,amssymb,nofootinbib,longbibliography]{revtex4-2}
\usepackage{graphicx} 
\usepackage{amsmath}
\usepackage{amsfonts,amsbsy}
\usepackage{amssymb}
\usepackage{relsize}
\allowdisplaybreaks

\usepackage{xcolor}
\definecolor{lcolor}{rgb}{0.5,0,0}
\definecolor{citcolor}{rgb}{0,0.3,0.0}
\usepackage[breaklinks,colorlinks,urlcolor=blue,citecolor=citcolor,linkcolor=lcolor]{hyperref}

\usepackage{mciteplus}

\def\gsim{ \,\, \vcenter{\hbox{$\buildrel{\displaystyle >}\over\sim$}}
 \,\,}

\def\be{\begin{equation}}
\def\ee{\end{equation}}
\def\bea{\begin{eqnarray}}
\def\eea{\end{eqnarray}}

\newcommand{\nc}{{N_\mathrm{c}}}
\newcommand{\dd}{{\rm d}}
\newcommand{\nn}{\nonumber}
\newcommand{\tr}{\, \mathrm{tr} }

\begin{document}

\title{Quark pair angular correlations in the proton:
entropy versus entanglement negativity}

\author{Adrian Dumitru}
\email{adrian.dumitru@baruch.cuny.edu}
\affiliation{Department of Natural Sciences, Baruch College, CUNY,
17 Lexington Avenue, New York, NY 10010, USA}
\affiliation{The Graduate School and University Center, The City University
  of New York, 365 Fifth Avenue, New York, NY 10016, USA}

\author{Eric Kolbusz}
\email{ekolbusz@gradcenter.cuny.edu}
\affiliation{Department of Natural Sciences, Baruch College, CUNY,
17 Lexington Avenue, New York, NY 10010, USA}
\affiliation{The Graduate School and University Center, The City University
  of New York, 365 Fifth Avenue, New York, NY 10016, USA}

\begin{abstract}
  Two-particle correlations in the proton on the light-front are
  described by a mixed density matrix obtained by tracing over all
  other, unobserved, degrees of freedom.  We quantify genuinely
  quantum quark azimuthal correlations in terms of the entanglement
  negativity measure of Quantum Information Theory.  While the
  two-quark state in color space is one of high entropy and weak
  quantum correlation, we find that a standard three-quark model wave
  function from the literature predicts an azimuthally correlated
  state of low entropy and high entanglement negativity. Low entropy
  is consistent with expectations for many colors (at fixed 't~Hooft
  coupling $g^2\nc$) but high negativity indicates substantial
  two-particle quantum correlations at $\nc=3$. We show that
  suppressing quantum correlations associated with entanglement
  negativity strongly modifies quark pair azimuthal moments $\langle
  \zeta^n\rangle$, $\zeta=
  \exp(i (\phi_1-\phi_2))$, intrinsic to the proton
  state.\\
  We also describe how to account for the leading ${\cal
    O}(g^2)$ correction to the density matrix from light-cone
  perturbation theory which is due to the presence (or exchange) of a
  gluon in the proton.  This correction increases the entropy and
  reduces the negativity of the density matrix for quark pair
  azimuthal correlations.  Hence, the entanglement negativity measure
  may provide novel insight into the structure of the proton state of
  QCD.
\end{abstract}

\maketitle
\tableofcontents
%\newpage

%-------------------------------------------------------------
\section{Introduction}

Entanglement, a quantum correlation in superposition states, is
generally regarded to be the most striking break of the quantum theory
of matter and radiation from classical theory~\cite{Schrodinger:1935q,
  Schrodinger:1936q, Einstein:1935q}.
In a light-front Fock state description of the proton, each Fock state
corresponds to a superposition of partons, quarks, anti-quarks, and gluons,
of all possible combinations of colors, flavors, spins, and momenta.
Entanglement of various degrees
of freedom in the proton is currently under intense
scrutiny~\cite{Kovner:2015hga, Kovner:2018rbf, Hagiwara:2017uaz,
  Kharzeev:2021nzh, Kharzeev:2017qzs, Beane:2019loz, Ehlers:2022oal,
  Tu:2019ouv, Kharzeev:2021yyf, Ramos:2020kaj, Hentschinski:2021aux,
  Hentschinski:2022rsa, Zhang:2021hra, H1:2020zpd, Armesto:2019mna,
  Duan:2020jkz, Duan:2021clk, Duan:2023zls, Dvali:2021ooc,
  Liu:2022hto, Liu:2022bru, Asadi:2022vbl, Asadi:2023bat,
  Dumitru:2022tud}.  One usually starts from the pure proton state and
traces over various unobserved degrees of freedom, the
``environment'', to obtain a reduced density matrix for the remaining
``system'':
\be
\rho_s = \tr_e\, \rho~.
\ee
In general, $\rho_s$ represents a mixed state. In this setting of
``bi-partite entanglement'' the magnitude of entanglement, i.e.\ of
quantum correlations, of the remaining degrees of freedom of the
system with those of the environment can be quantified, for example,
in terms of the von~Neumann entropy $S(\rho_s)$.  A pure state is
entangled if and only if the von~Neumann entropy of the partial state
$\rho_s$ is nonzero.\\

Our present focus is different. After tracing out the environment,
we divide further the remaining system into two systems $s_1$ and $s_2$.
We are interested in the sub-subsystem correlations of $s_1$ and $s_2$, 
specifically in azimuthal quark pair correlations in the proton, and 
whether these are quantum or classical, in the sense of Quantum 
Information Theory. The entanglement of $s_1$ with $s_2$ can not be 
measured via $S(\rho_{s_1})$ or $S(\rho_{s_2})$ because $\rho_s$ is not 
a pure state: these entropies are also sensitive to classical correlations
among the remaining two subsystems. Instead, we shall quantify the
magnitude of quantum correlations through the {\em entanglement
  negativity} of $\rho_s$~\cite{Vidal:2002q, Plenio:2005qn}.  A brief
introduction into separable states and quantum correlations can be
found in appendix~\ref{sec:QITbasics}.  \\

%-------------------------------------------------------------
\section{Color correlations}
\label{sec:qq_col}

We first present a simple yet instructive example.  We start from a
fully antisymmetric state of $\nc \ge 2$ color charges in the
fundamental representation of color-$SU(\nc)$:
\be
\rho_{i_1\ldots i_\nc, i'_1\ldots i'_\nc} =
\frac{1}{\nc!}\,
\epsilon^{i_1\ldots i_\nc}\epsilon^{i'_1\ldots i'_\nc}~.
\ee
Tracing over all but two degrees of freedom yields the 
two-subsystem reduced density matrix
\be \label{eq:rho-color}
\rho_{ij,i'j'}
= \frac{1}{\nc(\nc-1)}\left(\delta_{ii'}\delta_{jj'}
- \delta_{ij'}\delta_{i'j}\right)~.
\ee
For $N_c\to\infty$, at leading order in $1/N_c$ this reduces to
a product state
\be
\rho^\mathrm{LO}_{ij,i'j'}
= \frac{1}{\nc^2}\delta_{ii'}\delta_{jj'}~.
\ee
This state lacks correlations, and also is not in general
anti-symmetric under $i \leftrightarrow j$ (or $i' \leftrightarrow
j'$). The von~Neumann entropy for this matrix is $S^\mathrm{LO}=2\log
\nc$, {\em twice} the entropy for a single fundamental color charge,
i.e.\ the leading contribution to $S$ is extensive and scales with the
number of charges.

Correlations emerge at next-to-leading order,
\be \label{eq:rho-color-NLO}
\rho^\mathrm{NLO}_{ij,i'j'}
= \frac{1}{\nc^2}\left(\frac{\nc+1}{\nc}\delta_{ii'}\delta_{jj'} -
\delta_{ij'}\delta_{i'j}\right)~.
\ee
This density matrix does satisfy anti-symmetry, at leading order in
$1/\nc$.  Its two eigenvalues are $\lambda_1 = 1/\nc^3$
and $\lambda_2 = (2 + \frac{1}{\nc})/\nc^2$, with multiplicities
$N_1=\nc(\nc+1)/2$ and $N_2=\nc(\nc-1)/2$, respectively.
Hence, the entropy is $S^\mathrm{NLO} = 2\log\nc - \log 2 + \cdots$.  The
$\nc$-independent correction arises because the leading $2\log\nc$
overcounts the increase in the dimensionality of the Hilbert space
from one to two color charges.

Indeed, the two-particle Hilbert space ${\cal H} \otimes {\cal H} =
{\cal H}_S\oplus {\cal H}_A$ decomposes into a direct sum of a
symmetric and an anti-symmetric space, and the allowed state vectors
belong to the latter.  The dimension of ${\cal H}_A$ is
\be
{{\nc}\choose{2}} = \frac{\nc !}{2!\, (\nc-2)!} = \frac{1}{2}\nc (\nc-1)~,
\ee
since this is the number of linearly independent rank-2 anti-symmetric
tensors over a $\nc$-dimensional vector space.
Hence, this is the number of non-zero eigenvalues of the exact $\rho$ from
eq.~(\ref{eq:rho-color}). It is
clear that all eigenvalues are equal, so $\lambda_i=2/\nc(\nc-1)$.
This can be confirmed by explicit computation using standard techniques.

The purity of $\rho$ is $\tr\, \rho^2 = \sum_\lambda N_\lambda
\lambda^2 = \frac{2}{\nc (\nc-1)}$, and the entropy is $S=
-\sum_\lambda N_\lambda \lambda \log \lambda = \log [\frac{1}{2}\nc
  (\nc-1)]$. For $\nc=2$ the purity is $1$ and the entropy is $0$,
since nothing has been traced over and $\rho$ is a pure state. For
$\nc\to\infty$, on the other hand,
\be \label{eq:S-rho_col-large-Nc}
S(\rho) = 2\log\nc -\log 2 - \frac{1}{\nc} - \frac{1}{2\nc^2}
+ {\cal O}(\nc^{-3})~,
\ee
which exhibits subleading corrections due to quantum correlations.
Both terms, $-\log 2$ and $-\nc^{-1}$, are associated with the existence of a
negative eigenvalue of the partial transpose of $\rho$, as we discuss
at the end of this section (and in appendix~\ref{sec:PEN}).\\

Subsystem correlations may also be quantified in terms of the
``coherent information'' measure~\cite{Wilde:2011}. For a bipartite
state $\rho$ it is defined as
\be
I(2\rangle 1) \equiv S(\rho^{(1)}) - S(\rho)~,
\ee
where $\rho^{(1)} = \tr_2 \rho$ is the reduced density matrix for
system 1, and $S$ denotes the von~Neumann entropy. $I(2\rangle 1)$
quantifies how much less is known about subsystem 1 than about the
whole composed system~\cite{Wilde:2011}. In the presence of strong
entanglement and low entropy one expects $I(2\rangle 1)>0$ and vice
versa: if $\rho = \rho^{(1)}\otimes \rho^{(2)}$ then $S(\rho) =
S(\rho^{(1)}) + S(\rho^{(2)})$ and $I(2\rangle 1) = - S(\rho^{(2)})$.

For the density matrix~(\ref{eq:rho-color}),
$\rho^{(1)}_{ii'} = \nc^{-1}\delta_{ii'}$, and
\be \label{eq:I-rho-color}
I(2\rangle 1) = \log \frac{2}{\nc-1} = -\log\nc +\log 2
-\frac{1}{\nc}-\frac{1}{2\nc^2} + {\cal O}(\nc^{-3})~.
\ee
The leading contribution at large $\nc$ is, of course, half the
``ideal gas'' entropy: dividing the system in half reduces the entropy
by half. Hence, the negative coherent information indicates weak
entanglement and high entropy of the reduced state.
\\

The entanglement negativity is given by (minus) the sum of negative
eigenvalues of the partial transpose over the second system
$\rho^{T_2}$, which swaps $j$ and $j'$ in eq.~(\ref{eq:rho-color}).
The eigenvalues of $\rho^{T_2}$ are $-\frac{1}{\nc}$ with multiplicity
1 and $\frac{1}{\nc (\nc-1)}$ with multiplicity $\nc^2 - 1$. This
means that the negativity of $\rho$ is ${\cal N}(\rho)={1}/{\nc}$,
i.e.\ the inverse of the dimension of the Hilbert space for one
fundamental charge. Hence, quantum correlations besides
anti-symmetrization of the two remaining color charges are indeed
${\cal O}(\nc^{-1})$.  In the limit of many colors, this agrees with
the correlation entropy, i.e.\ the third term on the r.h.s.\ of
eqs.~(\ref{eq:S-rho_col-large-Nc}) or~(\ref{eq:I-rho-color}). However,
in this limit the overall entropy of the state~(\ref{eq:rho-color}) is
far greater than its negativity and coherent information is negative.

%-------------------------------------------------------------
\section{Light-cone wave functions and density matrices describing
  azimuthal correlations}
\label{sec:qq_angular}

We briefly introduce the light-cone Fock state description of the
proton state. Much more detailed accounts can be found in the
literature, e.g.\ refs.~\cite{Lepage:1980fj, Brodsky:1997de,
  Brodsky:2000ii, Brodsky:1994fz}.

A proton state with light-cone momentum $P^+$ and transverse momentum
$\vec P_\perp=0$ is written as
\be
|P\rangle = \sum_n \int \dd\Phi_n\, \Psi_n(k_1,\cdots,k_n)\,
|k_1,\cdots,k_n\rangle~.
\ee
We have omitted writing the spin-flavor and color space structure
since we will trace over those degrees of freedom. $\dd\Phi_n$
denotes the integration measure over the $n$ on-shell parton
three-momenta $k_i=(x_i P^+, \vec k_i)$, including $\delta$-functions
which enforce $\sum x_i=1$ and $\sum \vec k_i = \vec P_\perp=0$. The
amplitudes $\Psi_n(k_1,\cdots,k_n)$ are the $n$-parton light-cone wave
functions. They are gauge invariant and universal (process
independent), and are obtained, in principle, from the
non-perturbative solution of the QCD Hamiltonian. The ket
$|k_1,\cdots,k_n\rangle$ is obtained by acting with the appropriate
creation operators on the vacuum of the free theory, which in light-cone
quantization coincides with the vacuum of the interacting theory.\\

To date, exact solutions for the light-cone wave functions are not
available, of course. In the future, lattice gauge theory may provide
numerical solutions for moderate parton momentum fractions $x_i$ and
transverse momenta $\vec k_i$ via a large momentum expansion of
equal-time Euclidean correlation functions in instant
quantization~\cite{Ji:2020ect,Ji:2021znw,Liu-Zhao-Schafer-SNOWMASS21}.
In the following we shall rely on a truncation of Fock space and
solutions of effective light-cone Hamiltonians supplemented by the
${\cal O}(g^2)$ correction obtained from light-cone perturbation theory.

%------------------------------------------------
\subsection{Three quark Fock state}

Empirical observations suggest that at moderate momentum fractions
$x_i\gsim 0.1$, and for transverse momenta up to a few times the
QCD confinement scale, the light-cone momentum structure of the proton
is described reasonably well by a ``light front constituent quark
model''. In this approximation, the light-cone state of the proton is
written in terms of its three quark Fock state and an effective three
quark wave function as follows\footnote{Throughout the manuscript we
  write transverse momenta with and three-momenta without a vector
  arrow: $k=(xP^+,\vec k)$.}:
\bea
|P\rangle &=& \int\limits_{[0,1]^3} \prod_{i=1\cdots3}\frac{\dd x_i}{2x_i}
\, \delta\left(1-\sum_i x_i\right)
\int \prod_{i=1\cdots3}\frac{\dd^2 k_i}{(2\pi)^3}\,
(2\pi)^3\, \delta\left(\sum_i \vec k_i\right)\,
\Psi_\mathrm{qqq}\left(k_1; k_2; k_3\right)\,\,
\left|k_1; k_2; k_3\right> \nn\\
&=& 
\int \frac{\dd x_1 \dd x_2}{2x_1\, 2x_2\, 2(1-x_1-x_2)}
\int\frac{\dd^2 k_1}{(2\pi)^3}\,\frac{\dd^2 k_2}{(2\pi)^3}\,
\Psi_\mathrm{qqq}\left(x_1,\vec k_1; x_2, \vec k_2; 1-x_1-x_2, -\vec k_1
-\vec k_2\right) \nn\\
& & ~~~~~~~~~~~~
\left|x_1,\vec k_1; x_2,\vec k_2; 1-x_1-x_2,
-\vec k_1 -\vec k_2\right>~.
\label{Pstate_qqq}
\eea
As already mentioned above we omit the spin-flavor and color space
structures as we will focus on azimuthal correlations in momentum
space.  The spatial wave function $\Psi_\mathrm{qqq}$ is symmetric
under exchange of any two quarks: $\Psi_\mathrm{qqq}\left(k_1; k_2;
k_3\right) = \Psi_\mathrm{qqq}\left(k_2; k_1; k_3\right)$ etc.  The
second form of $|P\rangle$ shows that $k_3=(x_3 P^+,\vec k_3)$ is not
a degree of freedom, it has been eliminated by the COM constraint.  Only
$k_1$ and $k_2$ are degrees of freedom.

For numerical estimates below we employ a model due to Brodsky and
Schlumpf~\cite{Schlumpf:1992vq,Brodsky:1994fz} which we briefly
summarize for completeness.
Alternative models which represent solutions of effective light-cone
Hamiltonians with interactions can be found in the literature, e.g.\
refs.~\cite{Xu:2021wwj,Shuryak:2022thi}.

The model of Brodsky and Schlumpf used here corresponds to
\begin{equation}
  \Psi_\mathrm{qqq}\left(x_i,\vec k_i\right)
  = N  \, \sqrt{x_1 x_2 x_3}\,\,  e^{-{\cal M}^2/2\beta^2}
  \label{eq:Psiqqq_HO}
\end{equation}
where ${\cal M}^2 = \sum (\vec k_i^2+m_q^2)/x_i$ is the
invariant mass squared of the non-interacting three-quark
system~\cite{Bakker:1979eg}. It is understood that $x_3$ and $\vec
k_3$ are short-hands for $1-x_1-x_2$ and $-\vec k_1 - \vec k_2$,
respectively.  The normalization $N$ of this wave function follows from
$\tr\, \rho=1$, see below. The non-perturbative parameters
$m_q=0.26$~GeV and $\beta=0.55$~GeV have been tuned in
ref.~\cite{Brodsky:1994fz} to low-energy properties of the proton
such as its ``radius'' (the inverse RMS quark transverse momentum).\\

From the above expression for $|P\rangle$ one obtains the density matrix
\be \label{eq:rho_Psi*_Psi}
\rho_{\alpha \alpha'} =
\Psi_\mathrm{qqq}^*(k_1',k_2')\, \Psi_\mathrm{qqq}(k_1,k_2)~,
\ee
where $\alpha=\{k_1; k_2 \}$, $\alpha' =\{k_1' ; k_2' \}$; for a
detailed presentation of the steps from eq.~(\ref{Pstate_qqq})
to~(\ref{eq:rho_Psi*_Psi}) see ref.~\cite{Dumitru:2022tud}.
Here, we have omitted the momenta of the third quarks from the
arguments of the wave functions; they are understood to be such that
the sums of transverse momenta are zero while the sums of light-cone
momentum fractions are 1.

The trace measure is
\be \label{eq:tr_measure}
\tr = \frac{1}{2}\int \frac{\dd x_1 \dd x_2}{2x_1\, 2x_2\, 2(1-x_1-x_2)}
\int\frac{\dd^2 k_1}{(2\pi)^3}\,\frac{\dd^2 k_2}{(2\pi)^3}~,
\ee
and this sets the normalization of the light-cone wave function.

%--------------------------------
\subsection{Quark azimuthal angular correlations}
\label{sec:qq-azim-correl}

Our main interest in this paper is in two-quark angular
correlations. These are described by the density matrix $\rho_{\vec
  k_1\vec k_2,\vec k_1'\vec k_2'}$ obtained by tracing over $x_1$ and
$x_2$:
\be
\rho_{\vec k_1 \vec k_2, \vec k_1' \vec k_2'} = \int\frac{\dd x_1}{2x_1}
\frac{\dd x_2}{2x_2 (1-x_1-x_2)}\, \Psi^*(x_1,x_2,\vec k_1', \vec k_2')\,
\Psi(x_1,x_2,\vec k_1, \vec k_2)~.   \label{eq:rho_k1k2,k1'k2'-v1}
\ee
To reduce the dimension of the matrix we can also trace over
$|\vec k_1|$ and $|\vec k_2|$ to obtain
\be \label{eq:rho_phi1phi2}
\rho_{\phi_1\phi_2,\phi_1'\phi_2'} =
\int |\vec k_1|\, \frac{\dd |\vec k_1|}{16\pi^3}\, |\vec k_2|\,
\frac{\dd |\vec k_2|}{16\pi^3}\,
\int\frac{\dd x_1}{2x_1}
\frac{\dd x_2}{2x_2 (1-x_1-x_2)}\,
\Psi^*(k_1',k_2')\, \Psi(k_1,k_2)~.
\ee
Here, $\Psi^*$ involves the dot product $\vec k_1' \cdot \vec k_2' =
|\vec k_1|\, |\vec k_2|\cos(\phi_1'-\phi_2')$, and $\Psi$ involves
$\vec k_1 \cdot \vec k_2 = |\vec k_1|\, |\vec
k_2|\cos(\phi_1-\phi_2)$.  This means that the product $\Psi^*\, \Psi$
does not factorize into a function of $\phi_1, \phi_1'$ times a
function of $\phi_2, \phi_2'$. Hence, this is clearly not a product
state of the form $\rho^{(1)}\otimes\rho^{(2)}$. However, such a
product state emerges in the large-$\nc$ limit at fixed 't~Hooft
coupling $g^2\nc$ where the spatial wave function of $\nc$ quarks
factorizes into $N_c$ one-particle wave functions determined by a mean
field~\cite{Witten:1979kh}. The negativity of the corresponding
$\rho_{\phi_1\phi_2,\phi_1'\phi_2'}$ is zero.  \\

To actually construct this matrix on a computer we discretize the
angular interval $(-\pi,\pi]$ into a finite number of bins of size
  $\Delta\phi$. For proper normalization of the eigenvalues the
  r.h.s.\ of the previous expression should be multiplied by
  $(\Delta\phi)^2$. In particular, the trace will then be given simply
  by the sum of the diagonal elements of the matrix, as it should be.
  
We then determine numerically the eigenvalues of $\rho$ for various
bin sizes $\Delta\phi$ from $2\pi/16$ to $2\pi/128$. We find that the
entropy converges to $S(\rho)=0.25$. This occurs because as the number
of $\phi$ bins (and, hence, the number $N_\lambda$ of eigenvalues of
$\rho$) increases, the eigenvalue density of $\rho$ asymptotically 
approaches
\be \label{eq:dN/dlambda}
\frac{\dd N_\lambda}{\dd\lambda} = \left(N_\lambda-\sum_{i=1}^n C_i\right)\,
\delta(\lambda) +
\sum_{i=1}^n C_i\, \delta(\lambda-\lambda_i)~,
\ee
with $n$ the number of non-zero eigenvalues $\lambda_i$
with multiplicities $C_i$. Hence, $S=- \sum_{i=1}^n C_i \lambda_i \log
\lambda_i$.  Even lower entanglement entropies below 0.1 where obtained in
ref.~\cite{Dumitru:2022tud} for other spatial degrees of
freedom, using the same model light-cone wave function.

\begin{figure}[htb]
  \includegraphics[width=0.5\textwidth]{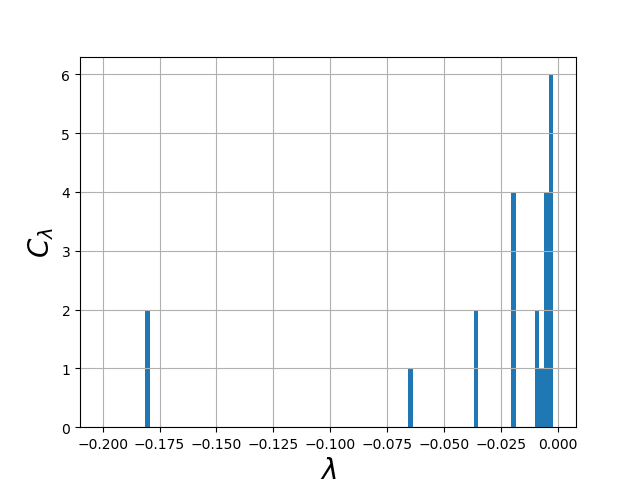}
  \vspace*{-.3cm}
  \caption{\label{fig:rhoT2-EV-density} Eigenvalue density of the
    partial transpose $\rho^{T_2}$ for 96 angular bins, plotted over $[-0.2,0.0)$.
      To suppress the $\delta(\lambda)$ peak the eigenvalue multiplicity in
      the first bin left of 0 is not shown.}
\end{figure}
The binned eigenvalue density of the partial transpose $\rho^{T_2}$
also converges to the form of eq.~(\ref{eq:dN/dlambda}), where now
some of the $\lambda_i$ are negative. The multiplicities $C_i$ of
negative eigenvalues are shown in fig.~\ref{fig:rhoT2-EV-density}.  We
obtain ${\cal N}(\rho)\simeq 0.68$.  This quantifies the magnitude of
two-quark azimuthal quantum correlations encoded in the
model wave function~(\ref{eq:Psiqqq_HO}), when all other degrees of
freedom have been traced over.  Contrary to the state in color space
described in sec.~\ref{sec:qq_col}, where the entropy is substantially
greater than the negativity (even for $\nc=3$ colors), here the
entropy of the angular density matrix is actually less than its
negativity. The coherent information measure described above confirms
the presence of genuine quantum correlations: for the angular
density matrix we obtain positive $I(2\rangle 1) = 0.32$. When
$\rho_{\phi_1\phi_2,\phi_1'\phi_2'}$ is subjected to ``classicalization''
via the PEN transformation described in appendix~\ref{sec:PEN},
coherent information turns negative: $I'(2\rangle 1) = -1.04$.
\\

To illustrate the potential relevance of entanglement negativity
to concrete observables we have computed the following
azimuthal angular moments of quark pairs in the proton:
\be
a_n \equiv \left< e^{i n (\phi_1-\phi_2)}\right>
= \int \dd\phi_1\dd\phi_2\, \rho_{\phi_1\phi_2,\phi_1\phi_2}\,
e^{i n (\phi_1-\phi_2)}
\ee
for $n=1, 2, 3$. As the diagonal of the density matrix is
symmetric under $\phi_1 \leftrightarrow\phi_2$ it follows that the
imaginary parts of the above moments vanish.
Table~\ref{tab:a_n} lists the values of $a_1, a_2, a_3$ obtained
with the three quark density matrix from eq.~(\ref{eq:rho_Psi*_Psi})
as well as with the modified $\rho'$ with vanishing entanglement
negativity; see appendix~\ref{sec:PEN} for a discussion of
the transformation $\rho\to\rho'$.
\begin{table}[hbt]
\begin{center}
\begin{tabular}{c|c|c}
$n$ & $a_n$ & $a_n'$ \\
\hline  
1 & $-0.404$ & $-0.121$ \\
2 & $0.152$ & $0.046$ \\
3 & $-0.058$ & $-0.017$
\end{tabular}
\caption{Angular moments $a_n = \left< e^{i n (\phi_1-\phi_2)}\right>$
  computed from the three quark density matrix. The primed
  moments correspond to the modified density matrix $\rho'$
  where quantum correlations associated with a negative eigenvalue
  of the partial transpose have been purged.
\label{tab:a_n}
}
\end{center}
\end{table}
We find substantial changes of the $a_n$ when the entanglement
negativity of the density matrix is erased. This is an indication that
quantum correlations intrinsic to the proton could be relevant for the
understanding of two-particle angular correlations in proton-nucleus
collisions~(see
refs.\cite{Kohara:2023qnp,Lappi:2015vta,Schenke:2015aqa} and
references therein) or deeply inelastic scattering (see below),
at least in the regime of moderately small $x$.

%---------------------------------------------------------
\section{Summary and Discussion}

In sec.~\ref{sec:qq_col} we consider the state of $\nc$ quarks in
color space. Tracing over the colors of $\nc-2$ quarks generates a
mixed state $\rho$ where the von~Neumann entropy $S(\rho) = \log
\frac{1}{2}\nc(\nc-1) = 2\log\nc -\log 2 -1/\nc + \cdots$ is much
greater than the level of quantum correlations measured by the
entanglement negativity, ${\cal N}(\rho) = 1/\nc$.
\\

In sec.~\ref{sec:qq_angular} we turn to our main focus, two-quark
azimuthal correlations in the proton on the light-front. We consider
moderate-energy scattering which probes parton fractional momenta
$x\gsim 0.1$, and transverse momenta not far beyond the QCD
confinement scale. In this regime, an effective description of the
proton in terms of a light-front constituent quark model should apply.
Tracing out all other degrees of freedom, we construct the reduced
density matrix $\rho_{\phi_1\phi_2,\phi_1'\phi_2'}$ which describes
angular correlations.  Using a standard light-cone model wave function
from the literature~\cite{Schlumpf:1992vq,Brodsky:1994fz} we make a
novel observation, that this state is characterized by low entropy,
$S\simeq 0.25$, and high entanglement negativity ${\cal
  N}\simeq0.68$. This is indicative of the presence of strong quantum
correlations among the two azimuthal angles, and of weaker
entanglement of the combined system $(\phi_1,\phi_2)$ with the traced
out ``environment''. The reduced state of the quark pair corresponds
to an entangled superposition of azimuths, not to a classical
statistical ensemble. For illustration we have computed $\langle\exp(
i n (\phi_1-\phi_2))\rangle$ moments of quark pair angular
correlations intrinsic to the proton from the three quark model
light-cone wave function. We find that these angular correlation
measures are clearly linked to the non-zero entanglement negativity
of the density matrix, i.e.\ to the presence of negative
eigenvalues of the partial transpose of $\rho$.\\

These azimuthal correlations could, in principle, be observed in
deeply-inelastic $e^- p$ scattering at the electron-ion collider
EIC~\cite{Accardi:2012qut, Aschenauer:2017jsk, Proceedings:2020eah,
  AbdulKhalek:2021gbh}.  In this process, a small quark anti-quark
dipole of transverse size $\vec r$ scatters from the proton at an
impact parameter $\vec b$, and the scattering amplitude $N(\vec r,
\vec b)$ depends on the azimuthal angle made by these two vectors.

Indeed, the angular dependence of
\begin{multline}
  N(\vec r,\vec b) = -g^4 C_F
  \int \frac{\dd^2 \vec K \dd^2 \vec q}{(2\pi)^4}
  \frac{\cos\left(\Vec b \cdot \vec K\right)}{(\vec q - \frac{1}{2}\vec K)^2\,\,
    (\vec q + \frac{1}{2} \vec K)^2}\,
  \left( \cos(\vec r \cdot \vec q) -
  \cos\left(\frac{\vec r \cdot \vec K}{2} \right) \!\! \right)
  \, G_2\left(\vec q -\frac{1}{2}\vec K, -\vec q - \frac{1}{2} \vec K\right)~,
  \label{eq:N_r_b}
\end{multline}
is determined by the angular dependence of the correlator of two
color charge density operators in the proton,
\be
\langle Q^a(\vec q_1)\, Q^b(\vec q_2) \rangle \equiv \delta ^{ab}\, g^2
G_2(\vec q_1,\vec q_2)~.
\ee
Restricting to the three quark Fock state for illustration, the result
for this correlator obtained in ref.~\cite{Dumitru:2018vpr} can be
rewritten in terms of the density matrix $\rho_{\alpha\alpha'}$
introduced above in eq.~(\ref{eq:rho_Psi*_Psi}):
\bea
G_2(\vec q_1,\vec q_2) &=&
\int\frac{\dd x_1\, \dd x_2}{8x_1 x_2 (1-x_1-x_2)}
\int\frac{\dd^2 k_1\, \dd^2 k_2}{(16\pi^3)^2} \left\{
\rho_{k_1 k_2, k_1' k_2'}-\rho_{k_1 k_2, \ell_1 \ell_2}\right\}~,
\label{eq:G2}
\eea
where $\vec k_1'=\vec k_1-(1-x_1)(\vec q_1+\vec q_2)$, $\vec k_2'=\vec
k_2+x_2(\vec q_1+\vec q_2)$, $\vec \ell_1 = \vec k_1-(1-x_1)\vec
q_1+x_1\vec q_2$, $\vec \ell_2 = \vec k_2-(1-x_2)\vec q_2+x_2\vec
q_1$, and $k_1^+ = k_1^{\prime +} = \ell_1^+ = x_1 P^+$, $k_2^+ = k_2^{\prime +}
= \ell_2^+ = x_2 P^+$.  The first and second terms of eq.~(\ref{eq:G2}) originate
from the ``handbag'' and ``cat's ears'' diagrams, respectively.
Note that this correlator satisfies a Ward identity and vanishes when
either $\vec q_1$ or $\vec q_2 \to 0$; this can be checked easily
using the permutation symmetry of the wave function.
Eqs.~(\ref{eq:N_r_b}, \ref{eq:G2}) describe the scattering of the dipole from
the entangled superposition state of the target.
\\

The
angular dependence of the correlator $G_2(\vec q_1,\vec q_2)$ and of
the dipole scattering amplitude $N(\vec r,\vec b)$ has been analyzed
in ref.~\cite{Dumitru:2021tvw}, and was shown to be qualitatively
different from ``geometry based'' models~\cite{Iancu:2017fzn}.
At smaller $x$, the angular dependence of $N(\vec r, \vec b)$ can,
alternatively, be attributed to the elliptic gluon Wigner
distribution~\cite{Hagiwara:2017ofm}.  The evolution of the azimuthal
dependence of $N(\vec r,\vec b)$ with $x$, in the quasi-classical
regime of very small $x$ has been analyzed in
refs.~\cite{Kovner:2010xk,Kovner:2011pe,Dumitru:2014vka}.\\

In appendix~\ref{sec:Og2}, we provide the expressions for the leading
${\cal O}(g^2)$ perturbative correction to the density matrix for
angular correlations. The additional presence (or the exchange) of a
gluon in the proton leads to a much wider range of parton light-cone
and transverse momenta.  Here,
$\rho^{(g^2)}_{\phi_1\phi_2,\phi_1'\phi_2'}$ is no longer a function
only of the differences $\phi_1-\phi_2$ and $\phi_1'-\phi_2'$, and so
the numerical cost of constructing the matrix increases by an order of
magnitude.  Nevertheless, to see how the perturbative correction
affects the entropy and entanglement negativity of the quark pair
density matrix we have performed a coarse numerical evaluation using
48 angular bins. We choose parameters so as to ensure that the
perturbative correction remains reasonably small,
i.e.\ $\alpha_s=0.1$, $\Delta^2=1$~GeV$^2$, $\Lambda^2=3$~GeV$^2$, and
$\langle x_q\rangle/x=3$. Even so, we obtain $S(\rho^{(g^2)}) \simeq
0.88$, and ${\cal N}(\rho^{(g^2)}) \simeq 0.58$.  Thus, as the longitudinal
and transverse phase space for the perturbative gluon opens up, there
is a substantial increase of the entropy and a slight drop of the
negativity.  We interpret this to indicate slightly weaker quark pair
azimuthal quantum correlations, and stronger entanglement of the
remaining azimuthal angles with the traced degrees of freedom.

%--------------------------------------------------
\section*{Acknowledgements}

We acknowledge support by the DOE Office of Nuclear Physics through
Grant DE-SC0002307, and The City University of New York for PSC-CUNY
Research grant 65079-00 53.

\appendix

%----------------------------------------------------------------------
\section{Separable vs.\ quantum correlated states}
\label{sec:QITbasics}

A product state is given by
\be  \label{eq:rho-product}
\rho = \sigma_1 \otimes \sigma_2~.
\ee
Here, $\sigma_1, \sigma_2$ are density matrices for subsystems 1
and 2, respectively; these may be pure or mixed states. Such a state
obviously describes uncorrelated subsystems. Also, the entropy is
additive, $S(\rho)=S(\sigma_1)+S(\sigma_2)$.

Now consider
\be \label{eq:cl-mix}
\rho = \sum_i p_i\,\, \rho^{(1)}_i \otimes \rho^{(2)}_i~.
\ee
This is a {\em classical statistical mixture} of product states
(enumerated by the index $i$) each with a probability weight $p_i$,
with $\sum_i p_i =1$. An example is given below for how such a state
may result from a partial trace over an entangled pure state. Also,
such states may be prepared through LOCC (local unitary operations and
classical communication) from a product state $\rho\otimes\sigma$:
\be
\rho\otimes\sigma \to \sum_i p_i \left( U_i\, \rho \, U^\dagger_i\right)
\otimes
\left( V_i\, \sigma\, V^\dagger_i\right)~.
\ee

In state (\ref{eq:cl-mix}), subsystems 1 and 2 do exhibit correlations:
given an observable $O = O^{(1)}\otimes O^{(2)}$ we have
\bea   \label{eq:cl-mix-<O>}
\tr \, O\rho = \sum_i p_i\,\,
\tr \left( O^{(1)}\rho^{(1)}_i\right)\,\,
\tr \left(O^{(2)}\rho^{(2)}_i\right)
&\neq&
\sum_i p_i\,\,
\tr \left( O^{(1)}\rho^{(1)}_i\right)\,\,
\sum_j p_j\,\,
\tr \left(O^{(2)}\rho^{(2)}_j\right) \\
& & = \tr\, \left( O^{(1)}\rho^{(1)}\right)\,
\tr\, \left( O^{(2)}\rho^{(2)}\right)
~.
\eea
Here, $\rho^{(1)}=\tr_2\,\rho=\sum_i p_i\, \rho^{(1)}_i$ denotes the
density matrix for subsystem 1, and $\rho^{(2)}$ that of subsystem 2.
In referring to~(\ref{eq:cl-mix}) as a classical mixture we do not
imply that either sub-system is in a classical state. These may well
be quantum states. However, we interpret the {\em correlations}
as classical since they are determined by the
classical probabilities $p_i$.
\\

A mixed density matrix that can not be written in the
form~(\ref{eq:cl-mix}) is said to exhibit quantum
correlations. Equivalently, $\langle O\rangle$ will not be given by a
convex sum of products of expectation values in the respective
subsystems, weighted by classical probabilities, like in
eq.~(\ref{eq:cl-mix-<O>}).
\\

Deciding whether a state $\rho$ is separable is called the
separability problem of Quantum Information Theory. It is believed to
be NP-hard in general~\cite{gurvits2002quantum,Horodecki:2009q}. One
available measure for quantum correlations is the so-called {\em
  negativity} ${\cal N}(\rho)$~\cite{Vidal:2002q}.  It is given by
minus the sum of negative eigenvalues of the ``partial transpose'' of
$\rho$ with respect to system~2~\cite{Peres:1996q,Horodecki:1996q}:
$\rho^{T_2} = (I\otimes T)(\rho)$, where $I$ and $T$ denote the
identity and transposition operators, respectively. Then,
\be
{\cal N}(\rho) = - \sum_{\lambda^{T_2}<0}\, \lambda^{T_2}~.
\ee
For a
state like eq.~(\ref{eq:cl-mix}) the negativity is zero since
\be
\rho^{T_2} = \sum_i p_i\,\, \rho^{(1)}_i \otimes \rho^{(2)\, T}_i
\ee
has the same eigenvalues as $\rho$ itself, all of which are $\ge
0$. Hence, negativity is ``blind'' to classical correlations, unlike
entanglement measures such as the von~Neumann entropy, which measure
both quantum and classical correlations. However, in high
dimensional Hilbert spaces negativity may vanish even when the state
does exhibit quantum correlations: ${\cal N}(\rho)=0$ is a necessary but not
a sufficient criterion for a given density matrix to be a separable
mixture like eq.~(\ref{eq:cl-mix}). Nevertheless, if ${\cal N}(\rho)>0$ then
$\rho$ is definitely not a sum of product states.
\\

A simple example for the emergence of a classical mixture of product
states from a partially traced pure state follows from a generalized
GHZ state~\cite{GHZ:2007q} of 3 or more qudits,
\be
\rho^\mathrm{GHZ} = \frac{1}{d} \sum_{i,j} |i,i,i\rangle\,
\langle j,j,j|~.
\ee
Here, $d$ is the dimension of the Hilbert space of each system.
Tracing out one of the qudits leaves a separable mixed state of the
form~(\ref{eq:cl-mix}),
\be
\tr_3\, \rho^\mathrm{GHZ} = \frac{1}{d} \sum_n |n,n\rangle\,
\langle n,n| = \frac{1}{d} \sum_n |n\rangle\,\langle n|\,
\otimes\, |n\rangle\, \langle n|~.
\ee
This density matrix has $d$ non-zero eigenvalues equal to $1/d$.  Its
entropy is $S=\log d$, and its negativity is ${\cal N}=0$. For $d\gg
1$ this is a high entropy classically correlated state without quantum
correlations. As a result of the strong classical correlations the
entropy is not proportional to the number of qudits left, i.e.\ it is
not extensive.

%----------------------------------------------------------------------
\section{PEN: purging quantum correlations associated with non-zero
negativity}
\label{sec:PEN}

Here we describe a transformation of a density matrix $\rho\to\rho'$
such that the negativity ${\cal N}(\rho')=0$, i.e.\ the partial
transpose of $\rho'$ does not have any negative eigenvalues.  If
$\rho$ represents a classical statistical mixture of product states as
in eq.~(\ref{eq:cl-mix}) then $\rho'=\rho$. In other words, classical
correlations associated with such a state are unaffected by the
transformation. We reiterate that ${\cal N}(\rho) = 0$ is a necessary
but not a sufficient criterion for a given density matrix to be a
separable mixture (except when the Peres-Horodecki criterion
applies). Hence, while the transformation we describe next eliminates
negative eigenvalues from the spectrum of the partial transpose, it
does not necessarily generate a separable state of the
form~(\ref{eq:cl-mix}).

We obtain $\rho'$ from the following sequence:
\be
\rho \to \sigma=\rho^{T_2} \to \sigma_D=U^\dagger\sigma U \to
\sigma_D' = \Theta(\sigma_D)\, \sigma_D \to
\sigma' = U\sigma_D' U^\dagger \to
\rho' = \frac{(\sigma')^{T_2}}{\mathrm{tr}\, \sigma'}~.
\label{eq:PEN}
\ee
That is, we diagonalize the partial transpose of $\rho$ through a
unitary transformation $U$ of the basis. We then
remove\footnote{Instead, one could also multiply the negative eigenvalues
  of $\sigma_D$ by a number $0<\zeta<1$ in order to reduce the negativity in
steps.} the negative eigenvalues of the partial transpose:
$\Theta(A)$ denotes a matrix valued Heavyside function which returns a
matrix with the same dimension as $A$ and with entries 0 or 1 if the
corresponding entry of $A$ is $\le0$ or $>0$, respectively.  We then
undo the basis rotation and the partial transposition, and rescale
$\rho'$ so that $\mathrm{tr}\, \rho' =1$.  Note that $\rho'$ is
positive semi-definite and hermitian (if $\rho$ is), and so it
represents a valid density matrix.
Since $\rho' \ne \rho$ if ${\cal N}(\rho)>0$, through this transformation
one may study how subsystem correlations $\mathrm{tr}\, O^{(1)}\otimes O^{(2)}
\, \rho$ are modified by the ``Purge Entanglement Negativity''
(PEN) transformation.
\\

When $\sigma=\rho^{T_2}$ has one single negative eigenvalue
$\lambda_-$, and a degenerate spectrum of $n-1$ positive eigenvalues
$\lambda_+$ then the above transformation corresponds to shifting
$\lambda_-$ up to 0, and shifting all positive eigenvalues down by
$\lambda_-/(n-1)$. That is, in this case $\rho'$ is obtained from
$\rho$ by adding the following traceless matrix, in order to maintain
normalization:
\be \label{eq:rho'-rho-shift}
\rho' = \rho + a (1\!\!1 - n\rho)~.
\ee
The transformation~(\ref{eq:PEN}) corresponds to choosing the minimal
value for $a$ that leads to ${\cal N}(\rho')=0$.
\\

We now provide an explicit example for the color space density
matrix given previously in eq.~(\ref{eq:rho-color}):
\be \label{eq:rho-color-b}
\rho_{ij,i'j'}
= \frac{1}{\nc(\nc-1)}\left(\delta_{ii'}\delta_{jj'}
- \delta_{ij'}\delta_{i'j}\right)~.
\ee
From here onward we consider $\nc\ge3$ since ${\cal N}(\rho)=0$
for $\nc=2$, so $\rho'=\rho$ in that case.

The PEN transformation~(\ref{eq:PEN}) results in
\be \label{eq:rho-color-PEN}
\rho' = \frac{1}{\nc (\nc+1)} \, 1\!\!1 + \frac{1}{\nc+1}\, \rho~.
\ee
On the other hand, to determine the allowed values for $a$ from
eq.~(\ref{eq:rho'-rho-shift}) we first write the explicit form
of the partial transpose of $\rho$ as a $\nc^2\times\nc^2$ matrix:
\be
\renewcommand\arraystretch{1.2}
\sigma = \rho^{T_2} = \frac{1}{\nc(\nc-1)}\begin{bmatrix}
0\vspace{0.25em} & \overbrace{0 \cdots 0}^{\nc \rm\ times} & -1 & \overbrace{0 \cdots 0}^{\nc \rm\ times} & -1 & \cdots & -1 &  \overbrace{0 \cdots 0}^{\nc \rm\ times} & -1 \\

\shortstack{0\vspace{-0.5em}\\\vdots\\0} & \shortstack{\ \vspace{-0.5em}\\I\\{\color{white} 0}} & \shortstack{0\vspace{-0.5em}\\\vdots\\0} & \shortstack{\ \vspace{-0.5em}\\0\\{\color{white} 0}} & \shortstack{0\vspace{-0.5em}\\\vdots\\0} &  & \shortstack{0\vspace{-0.5em}\\\vdots\\0} & \shortstack{\ \vspace{-0.5em}\\0\\{\color{white} 0}} & \shortstack{0\vspace{-0.5em}\\\vdots\\0} \\

-1\vspace{0.25em} & 0 \cdots 0 & 0 & 0 \cdots 0 & -1 & \cdots & -1 & 0 \cdots 0 & -1 \\

\shortstack{0\vspace{-0.5em}\\\vdots\\0} & \shortstack{\ \vspace{-0.5em}\\0\\{\color{white} 0}} & \shortstack{0\vspace{-0.5em}\\\vdots\\0} & \shortstack{\ \vspace{-0.5em}\\I\\{\color{white} 0}} & \shortstack{0\vspace{-0.5em}\\\vdots\\0} &  & \shortstack{0\vspace{-0.5em}\\\vdots\\0} & \shortstack{\ \vspace{-0.5em}\\0\\{\color{white} 0}} & \shortstack{0\vspace{-0.5em}\\\vdots\\0} \\

-1\vspace{0.25em} & 0 \cdots 0 & -1 & 0 \cdots 0 & 0 & \cdots & -1 & 0 \cdots 0 & -1 \\

\vdots\vspace{0.4em} & & \vdots && \vdots & \ddots & \vdots & & \vdots\\

-1\vspace{0.4em} & 0 \cdots 0 & -1 & 0 \cdots 0 & -1 & \cdots & -1 & 0 \cdots 0 & 0 \\

\shortstack{0\vspace{-0.5em}\\\vdots\\0} & \shortstack{\ \vspace{-0.5em}\\0\\{\color{white} 0}} & \shortstack{0\vspace{-0.5em}\\\vdots\\0} & \shortstack{\ \vspace{-0.5em}\\0\\{\color{white} 0}} & \shortstack{0\vspace{-0.5em}\\\vdots\\0} &  & \shortstack{0\vspace{-0.5em}\\\vdots\\0} & \shortstack{\ \vspace{-0.5em}\\I\\{\color{white} 0}} & \shortstack{0\vspace{-0.5em}\\\vdots\\0} \\

-1 & 0 \cdots 0 & -1 & 0 \cdots 0 & -1 & \cdots & -1 & 0 \cdots 0 & 0
\end{bmatrix}\ . \label{eq:rho-color-ppt-general}
\ee
The eigenspaces of $\sigma$ decouple since the $\nc - 1$ identity 
matrices have no overlap with the grid of $-1$s; trivially there are 
$(\nc-1)\nc$ eigenvectors with eigenvalue $[\nc(\nc-1)]^{-1}$. Adding 
a diagonal matrix preserves the eigenspaces, so we can easily find the 
eigenvalues of $\sigma'$ by taking the partial transpose of 
\eqref{eq:rho'-rho-shift}. The identity blocks on the diagonal of 
\eqref{eq:rho-color-ppt-general} are all multiplied by $(1-a\nc^2)$, 
then adding $aI$ means that the new eigenvalues are
\be
a + (1-a\nc^2)\cdot \frac{1}{\nc(\nc-1)} = 
\frac{1}{\nc-1}\left(\frac{1}{\nc} - a \right)\ \label{eq:colorlambdaplus}.
\ee

The remaining eigenvalues are the eigenvalues of the $\nc$-by-$\nc$ matrix
\be
\begin{bmatrix}
a & y & y & & y \\
y & a & y & \cdots & y \\
y & y & a & & y \\
  & \vdots & & \ddots & \vdots \\
y & y & y & \cdots & a
\end{bmatrix} \label{eq:rho-color-subspace-general}
\ee
where
\be
y = (1-a\nc^2)\cdot\frac{-1}{\nc(\nc-1)} = \frac{a\nc^2-1}{\nc(\nc-1)}\ .
\ee
One eigenvector of \eqref{eq:rho-color-subspace-general} is 
$\sum_i \hat{e}_i$ with eigenvalue $a + (\nc-1)y$. The remaining 
$\nc-1$ eigenvectors are of the form $\hat{e}_1 - \hat{e}_i$ 
(for all $i \neq 1$) and all have eigenvalue $a-y$, which when worked 
out is equal to \eqref{eq:colorlambdaplus}. Therefore the eigenvalues 
of $\sigma'$ in \eqref{eq:rho'-rho-shift} are $\lambda_1$ with 
multiplicity $\nc^2-1$ and $\lambda_2$ with multiplicity 1, where
\bea
\lambda_1 &= \frac{1}{\nc-1}\left(\frac{1}{\nc} - a \right) \\
\lambda_2 &= a(\nc + 1) - \frac{1}{\nc}~.
\eea
If both of these eigenvalues are to be non-negative as to make 
${\cal N}(\rho')=0$, we require
\be
\frac{1}{\nc(\nc+1)} \leq a \leq \frac{1}{\nc}\ .
\ee

For the minimal $a=[\nc (\nc+1)]^{-1}$, the spectrum of
$\rho'$ as given in eq.~(\ref{eq:rho-color-PEN}) is
$\lambda=(\nc^{2}-1)^{-1}$ with multiplicity $\nc^2-1$, and
$\lambda=0$ with multiplicity 1. Hence, after removal
of the negative eigenvalue of $\rho^{T_2}$ the entropy increases to
\be
S(\rho') = \log(\nc^2-1) = 2\log\nc - \frac{1}{\nc^2} + \cdots
\ee
Comparing to eq.~(\ref{eq:S-rho_col-large-Nc}) for $S(\rho)$ we note
that the minimal shift $\rho\to\rho'$ has removed the two leading
correlation contributions: $-\log 2$ from anti-symmetrization as well
as the term $-1/\nc$, which is (minus) the negativity. Some correlation
contributions to the entropy at integer powers of $\nc^{-2}$ are
still present, however.

%----------------------------------------------------------------------
\section{${\cal O}(g^2)$ correction to the angular density matrix}
\label{sec:Og2}

In this section we list the leading ${\cal O}(g^2)$ correction to the
angular density matrix.  The details of the calculation of the one
gluon emission/exchange correction to the proton state $|P\rangle$ in
light-cone perturbation theory have been published in
refs.~\cite{Dumitru:2022tud, Dumitru:2020gla}. Here, we restrict to
quoting the resulting expressions for the angular density matrix.

The density matrix is now given by the original LO quark density
matrix plus the ${\cal O}(g^2)$ virtual correction(s), plus the
four-particle (qqqg) density matrix traced over the gluon.

We will restrict to the limit where the light-cone momentum cutoff $x$
for the gluon is much less than typical quark light-cone momentum
fractions.  Although not strictly required this kinematic restriction
greatly simplifies the following expressions.  \\

The first virtual correction arises when a quark in $|P\rangle$ or
$\langle P|$ exchanges a gluon with itself. This replaces $\Psi^*
\Psi$ in eq.~(\ref{eq:rho_Psi*_Psi}, \ref{eq:rho_phi1phi2}) by
\be \label{eq:rho-qq_virt1}
\Psi_\mathrm{qqq}^*(k_1',k_2')\, \Psi_\mathrm{qqq}(k_1,k_2)\,
\left[1 - 3\, C_q(\langle x_q\rangle;\, x,{\Lambda}/{\Delta})
  \right]
\ee
where
\bea
C_q(\langle x_q\rangle;\, x,\Lambda/\Delta) &=&
4 g^2 C_F \int\limits_x^{\langle x_q\rangle}
\frac{\dd x_g}{x_g}\frac{\dd^2 k_g}{16\pi^3}\,
\left[\frac{1}{k_g^2+\Delta^2} - \frac{1}{k_g^2+\Lambda^2}\right] \\
&=&
\frac{4 g^2 C_F}{16\pi^2} \, \log\frac{\langle x_q\rangle}{x}\,
\log\frac{\Lambda^2}{\Delta^2}~.
\eea
Here, $C_F=4/3$ is the eigenvalue of the quadratic Casimir in the
fundamental representation of color-$SU(3)$, and $\Delta$ and
$\Lambda$ denote a collinear regulator and a UV subtraction point,
respectively. For simplicity, here we take both to be constants,
independent of $x_g$.

The second virtual correction is due to the exchange of a gluon by two
quarks in $|P\rangle$ or $\langle P|$. When quarks 1 and 2 in $|P\rangle$
exchange a gluon,
\bea
\rho^{(12)}_{\alpha\alpha'} &=& \frac{2 g^2 C_F N_c}{3}\int_x\frac{\dd x_g}{x_g}
\frac{\dd^2k_g}{16\pi^3} \frac{1}{k_g^2+\Delta^2}\,
\Psi^*(k_1',k_2')\, \Psi(k_1+k_g,k_2-k_g)~.
 \label{eq:rho-qq_virt2}
\eea
We add this to eq.~(\ref{eq:rho-qq_virt1}). In fact, we need to also add
the virtual corrections from gluon exchanges by other quarks, so we
replace $\Psi^*(k_1',k_2')\, \Psi(k_1+k_g,k_2-k_g)$ in the previous expression
by
\bea
& &
\Psi^*(k_1', k_2')\, \Psi(k_1+k_g,k_2-k_g) +
\Psi^*(k_1', k_2')\, \Psi(k_1+k_g,k_2) +
\Psi^*(k_1', k_2')\, \Psi(k_1,k_2+k_g) \nn\\
&+&
\Psi^*(k_1'+k_g,k_2'-k_g)\, \Psi(k_1, k_2) +
\Psi^*(k_1'+k_g,k_2')\, \Psi(k_1, k_2) +
\Psi^*(k_1',k_2'+k_g)\, \Psi(k_1, k_2) ~.
 \label{eq:rho-qq_virt3}
\eea
\\

We continue with the real emissions. Consider first the case where a
gluon is emitted from quark 1 in $|P\rangle$ and quark 1' in $\langle
P|$. Eq.~(51) of ref.~\cite{Dumitru:2022tud} leads to the following
expression for the four-particle qqqg state:
\be
\rho_{\alpha\alpha'}^{(11')} = \frac{4 g^2 C_F N_c}{3}\,
\Psi^*(k_1', k_2')\, \Psi(k_1, k_2)\,
\frac{\vec k_g\cdot\vec k_g'}{(k_g^2+\Delta^2)\,
    (k_g^{\prime 2}+\Delta^{2})}~.
\ee
The trace over quark and gluon colors has already been performed, and
the limit $x_g\ll x_1$, $x_g'\ll x_1'$ has been taken.  The matrix
indices are $\alpha=\{x_1,\vec k_1, x_2, \vec k_2, x_g, \vec k_g\}$
and $\alpha'=\{x_1,\vec k_1', x_2, \vec k_2', x_g', \vec k_g'\}$.

Here we have to mind a subtlety: the above density matrix was
obtained by projecting $|P\rangle$ onto $\langle\alpha| = \langle
k_1-k_g; k_2; k_3|$ and $\langle P|$ onto $|\alpha'\rangle =
|k_1'-k_g'; k_2'; k_3'\rangle$, with $\sum\vec k_i = \sum\vec k_i' = 0$.
When we trace over the gluon we want to keep the quark momenta fixed,
however. Hence, we need to first shift $k_1\to k_1+k_g$ and $k_1'\to k_1'+k_g'$,
and only then do we trace over $k_g$. This leads to
\be  \label{eq:real1}
\rho_{\alpha\alpha'}^{(11')} = \frac{4 g^2 C_F N_c}{3}\,
\int\frac{\dd x_g}{x_g}\frac{\dd^2k_g}{16\pi^3}
\Psi^*(k_1'+k_g,k_2')\, \Psi(k_1+k_g,k_2)\,
\left[\frac{1}{k_g^2+\Delta^2} - \frac{1}{k_g^2+\Lambda^2}\right]~.
\ee
Here the UV subtraction has been included so that the trace is
finite. Now the matrix indices are $\alpha=\{x_1,\vec k_1, x_2, \vec k_2\}$ and
$\alpha'=\{x_1,\vec k_1', x_2, \vec k_2'\}$.
The sums over the transverse momentum arguments of the wave
functions are still 0, so $\Psi^*$, $\Psi$ are evaluated for $\vec
k_3'=-(\vec k_1'+\vec k_2'+\vec k_g)$ and $\vec k_3=-(\vec k_1+\vec
k_2+\vec k_g)$, respectively.  On the other hand, $x_3$ and $x_3'$ are
still given by $1-x_1-x_2$ and $1-x_1'-x_2'$, respectively, because we
assumed that the integral is dominated by negligibly small $x_g$.
This expression, plus analogous contributions which account for gluon
emissions from 2, 2' and 3, 3', now have to be added to the integrand
of eq.~(\ref{eq:rho_Psi*_Psi}).\\ (Cross check: if we trace over
quarks with the measure~(\ref{eq:tr_measure}) then we can shift $k_1$
back, and once we multiply by 3 to also account for gluon emissions
from 2, 2' and 3, 3', then we reproduce eq.~(71) of
ref.~\cite{Dumitru:2022tud}. Furthermore, this contribution then
cancels exactly against the ${\cal O}(g^2)$ correction from
eq.~(\ref{eq:rho-qq_virt1}), as it should to preserve the trace of the
density matrix.)\\

The second real emission correction is due to two different quarks
in $|P\rangle$ and $\langle P|$ each emitting a gluon. Eq.~(56) of
\cite{Dumitru:2022tud}, traced over quark colors, gives the four particle
state
\be
\rho_{\alpha\alpha'}^{(12')} = - \frac{2 g^2 C_F N_c}{3}\,
\Psi^*(k_1', k_2')\, \Psi(k_1, k_2)\,
\frac{\vec k_g\cdot\vec k_g'}{(k_g^2+\Delta^2)\,
    (k_g^{\prime 2}+\Delta^2)}~.
\ee
Here we need to shift $k_1\to k_1+k_g$ and $k_2'\to k_2'+k_g'$, and then
we can trace out the gluon:
\be  \label{eq:real2}
\rho_{\alpha\alpha'}^{(12')} = - \frac{2 g^2 C_F N_c}{3}\,
\int\frac{\dd x_g}{x_g}\frac{\dd^2k_g}{16\pi^3}
\Psi^*(k_1',k_2'+k_g)\, \Psi(k_1+k_g,k_2)\,
\frac{1}{k_g^2+\Delta^2}~.
\ee
Once again, here the transverse components of $k_3, k_3'$ are such
that the sum of transverse momenta is 0. If we trace out the
quarks, too, then $k_1'=k_1$ and $k_2'=k_2$; we are then allowed to shift
$k_2\to k_2-k_g$ and (\ref{eq:real2}) cancels against~(\ref{eq:rho-qq_virt2})
so that the proper normalization of the density matrix is
preserved.\\

In summary, to account for the ${\cal O}(g^2)$ perturbative correction,
in eq.~(\ref{eq:rho_phi1phi2}) for
$\rho_{\phi_1\phi_2,\phi_1'\phi_2'}$ one replaces the LO density
matrix $\Psi^*()\, \Psi()$ by the sum of eqs.~(\ref{eq:rho-qq_virt1}),
(\ref{eq:rho-qq_virt2},\ref{eq:rho-qq_virt3}), (\ref{eq:real1}) plus
analogous contributions for (11') $\to$ (22'), (33'),
and~(\ref{eq:real2}) plus analogous contributions for (12') $\to$
(13'), (21'), (23'), (31'), (32').
As we have indicated, the perturbative correction cancels in the
sum of eigenvalues of the density matrix, which is therefore independent
of the perturbative coupling $g^2$, the gluon light-cone momentum cutoff $x$,
the collinear regulator $\Delta$, and the UV regulator $\Lambda$.
However, the {\em spectrum} of eigenvalues does depend on these
quantities, and so does the entropy and the negativity of $\rho$.

\bibliography{refs}

\end{document}